\documentstyle[12pt]{article}
\newcommand{\eqref}[1]{(\ref{#1})}
\newcommand{\cL}{{\cal L}}

\renewcommand{\d}{\partial}

\newcommand{\nn}{\nonumber\\}

\newcommand{\ph}{\varphi}

\newcommand{\bP}{\bar\Phi}
\newcommand{\exv}[1]{\left\langle{#1}\right\rangle}

\newcommand{\ep}{\varepsilon}

\newcommand{\q}{{\bf q}}

\newcommand{\p}{{\bf p}}

\newcommand{\pint}[2]{{\int\!\frac{d^{#1}#2}{(2\pi)^#1}\,}}

\begin{document}
\pagestyle{empty}
\begin{flushright}
{CERN-TH/2001-074}\\
{ITP-Budapest 564}\\
hep-ph/0103093 \\
\end{flushright}
\vspace*{5mm}
\begin{center}
{\bf EFFECT OF THE SCALAR CONDENSATE\\ 
  ON THE LINEAR GAUGE FIELD RESPONSE\\ 
  IN THE ABELIAN HIGGS MODEL}\\
\vspace*{0.8cm}
 
A. Jakov\'ac$^{a,}$\footnote{e-mail: Antal.Jakovac@cern.ch},
A. Patk{\'o}s$^{b,}$\footnote{e-mail: patkos@ludens.elte.hu},
Zs. Sz{\'e}p$^{b,}$\footnote{e-mail: szepzs@cleopatra.elte.hu}\\

\vspace{0.3cm}
{\em 
$^{\rm a}$ Theory Division, CERN, CH-1211 Geneva 23, Switzerland\\
$^{\rm b}$ Department of Atomic Physics, E{\"o}tv{\"o}s University,
H-1117 Budapest, Hungary}

\vspace*{2cm}  
{\bf ABSTRACT} \\ \end{center}
\vspace*{5mm}

\noindent
The effective equations of motion for low-frequency mean gauge fields
in the Abelian Higgs model are investigated in the presence of a
scalar condensate, near the high temperature equilibrium.  We
determine the current induced by an inhomogeneous background gauge
field in the linear response approximation up to ${\cal O}(e^4)$,
assuming adiabatic variation of the scalar fields.  The physical
degrees of freedom are found and a physical gauge choice for the
numerical study of the combined Higgs+gauge evolution is proposed.

\vspace*{0.5cm}
\begin{flushleft} CERN-TH/2001-074 \\ITP-BUDAPEST 564\\
March 2001
\end{flushleft}
\vfill\eject

\setcounter{page}{1}
\pagestyle{plain}

\section{Introduction}
The real-time transformation of light gauge fields
into massive intermediate vector bosons is a subject of increasing
cosmological interest. 

Recently, Garcia-Bellido {\it et al.} \cite{Garcia99} (see also
\cite{Krauss99}) assumed that the coupled inflaton+Higgs system was
far out of equilibrium when the energy density of the expanding
Universe has passed the point corresponding to the thermal Higgs
transition. With the supplementary condition that the reheating
temperature after the exit from the inflationary period did not exceed
the electroweak critical temperature, they demonstrated in a
(1+1)-dimensional toy model, that large enough matter--antimatter
asymmetry could have survived till today. The existence of
non-equilibrium Higgs transitions in 3+1 dimensions has been
demonstrated by Rajantie {\it et al.} \cite{Rajantie00}.  In these
investigations the classical mean field equations were used with
renormalized and temperature-dependent couplings.

A more traditional field of the investigation is the evolution of the
baryon asymmetry through the electroweak phase transition,
accompanying the onset of the Higgs effect. It has been thoroughly
discussed, under the assumption that the system is in thermal
equilibrium, with specific emphasis on the high-temperature sphaleron
rate \cite{Moore01,Moore00,Ambjorn95}. Near equilibrium, the
gauge-invariant HTL action dominates the influence of the thermalized
quantum fluctuations on the motion of the mean fields with
$k\simeq{\cal O}(gT)$ \cite{Braaten90,Frenkel90,Blaizot94}. This term
was taken into account in real-time simulations of the sphaleron rate
\cite{Moore00} (see also \cite{Hindmarsh99}).

The framework for the theoretical study of the baryon asymmetry has
been somewhat modified with the realization that, within the Standard
Model, the Higgs effect sets in via smooth phase transformation,
characterized by an analytic variation of the order parameter
\cite{Kajantie97,Karsch97,Guertler97,Csikor99}.  Under such
circumstances the expectation value of the Higgs field is different
from zero also above the electroweak energy scale.  In this context it
is remarkable that the leading HTL correction is insensitive to the
presence of the scalar condensate \cite{Rebhan95, Moore00}.

In this note we go one step beyond the derivation of the HTL
correction. Our aim is to find the leading effect due to the presence
of an arbitrary constant background field: $\overline{\Phi}$. In
equilibrium, such a background is sufficient for the calculation of
the effective potential, but not the full effective action. Similarly
here, we restrict somewhat the generality of our correction terms to
the equations of motion, namely the corrections to the $\Phi^2$--$A^2$
vertex will be determined neglecting the non-local effects in the
scalar field.

The high-temperature limit of the resulting expression of the induced
current shows that the next-to-leading ``mass and vertex'' corrections
can be uniquely decomposed into gauge-fixing invariant and
gauge-fixing dependent modes.  In this way we can propose an
expression of wider applicability for the Landau damping effect in the
presence of a scalar condensate.

Baacke and his collaborators \cite{Baacke97,Baacke99,Heitmann01} have
applied, in a series of papers, a complementary approach to the Higgs
effective action without assuming the presence of a general gauge
background.

A non-trivial mean gauge field configuration reacts back on the scalar
condensate. Our final goal is to derive a coupled set of equations for
the scalar condensate and the mean gauge field, which is appropriate
for studying the real-time onset of the Higgs effect and the variation
of the sphaleron generation and decay rate. We propose a physical
gauge for the numerical solution of these equations where the
gauge-fixing dependent mode does not propagate.

The generalization of the derivation to non-Abelian systems presents
only technical complications.  The Abelian calculation has been
performed both in the standard real-time generator functional
formulation \cite{LeBellac96} and with an iterated solution of the
linearized Heisenberg equations of the quantum fluctuations
\cite{Boyanovsky98,Heinz86,Mrovczinsky90}. We shall use below the
latter approach, where the adequate two-point functions enter more
naturally. The general linear response theory will be described in a
separate publication \cite{Jakovac01}.

\section{Mean Field Equations in the Abelian Higgs model}

The Lagrangian of the model in the $O(2)$ notation is the following:
\begin{equation}
  \cL = -\frac14 \hat F_{\mu\nu}\hat F^{\mu\nu} +\frac12 (\d_\mu\hat
  \Phi)^2 - \frac12 m^2\hat \Phi^2 + e\hat A_\mu \hat J^\mu +\frac{e^2}2
  \hat A^2\hat \Phi^2 -\frac\lambda{24}(\hat \Phi^2)^2,
\end{equation}
where $\hat \Phi=(\hat \Phi_1,\hat \Phi_2)$ and $\hat J_\mu = \hat
\Phi_2\d_\mu\hat \Phi_1 - \hat \Phi_1\d_\mu\hat \Phi_2$. We split the
fields into a mean field ($A_\mu,\Phi$) and a fluctuation contribution
\begin{equation}
  \hat A_\mu =A_\mu+a_\mu,\quad \hat \Phi=\Phi+\ph,\quad
\exv{a_\mu}=\exv{\ph}=0
\end{equation}
at any time (averaging is understood with respect to the initial
density matrix). We assume that the scalar background is constant and
that it points to the $\Phi_1$ direction, its value being $\bP$. We
use in the sequel the notations $m_W=e\bP,\, m_H^2=m^2+\lambda
\bP^2/2,\, m_G^2=m^2+\lambda\bP^2/6$, although these are not the
vacuum values.

We fix a 't Hooft $R_\xi$ gauge by changing the Lagrangian as
\begin{equation}
  \cL\to\cL -\frac1{2\xi}(\d_\mu \hat A^\mu + \xi m_W\ph_2)^2 -\bar
  c(\d^2+\xi m_W^2 +e\xi m_W\ph_1)c,
\label{gaugeEOM}
\end{equation}
where $c$ is the ghost field. In this way the field $\ph_2$ receives
an additional mass contribution $\xi m_W^2$.

The operatorial equation of motion (EOM) separately induces EOMs for
the mean fields and the fluctuations. The average of the operatorial
EOM for the gauge field reads as
\begin{equation}
  \left[(\d^2+m_W^2)g_{\mu\nu} - \left(1-\frac1\xi\right) \d_\mu\d_\nu
  \right] A^\nu + j^{ind}_\mu(A) =0,
\label{gauge_eq}
\end{equation}
with the induced current
\begin{equation}
  j^{ind}_\mu(A) = e\exv{j_\mu} + 2e^2\bP\exv{a_\mu\ph_1} +
  e^2A_\mu\exv{\ph^2} + e^2\exv{a_\mu \ph^2},
\label{ind_curr}
\end{equation}
where $j_\mu=\ph_2\d_\mu\ph_1-\ph_1\d_\mu\ph_2$. We will perform the
calculations at the one-loop level, when the last term does not
contribute. The subtraction of (\ref{gauge_eq}) from the full equation
yields the EOM for the fluctuations. At one loop it is sufficient to
consider only the equations linearized in the fluctuations.  On the
other hand, since we want to calculate $j^{ind}_\mu$ in the linear
response approximation, we can neglect all terms non-linear in $A$.
These assumptions make the equations very simple:
\begin{eqnarray}
  && \left[-(\d^2+m_W^2)g_{\mu\nu} +\left(1-\frac1\xi\right)
    \d_\mu\d_\nu \right] a^\nu - 2e^2\bP A_\mu \ph_1 = 0,\nn
  && (\d^2+m_H^2) \ph_1 + 2eA^\mu\d_\mu\ph_2 +e(\d A)\ph_2 -2e^2 \bP
  A^\mu a_\mu =0,\nn
  && (\d^2+m_G^2+\xi m_W^2) \ph_2 - 2eA^\mu\d_\mu\ph_1 - e(\d A)\ph_1=0.
\label{linopeom}
\end{eqnarray}
The ghost fields follow a free EOM, so that they do not influence the
present calculation. These equations are solved to linear order in the
$A$ background:
\begin{eqnarray}
  a_\mu(x) && = a_\mu^{(0)}(x) - 2e^2\bP \int\! d^4z\,
  G^R_{\mu\nu}(x-z) A^\nu(z) \ph_1^{(0)}(z),\nn
  \ph_1(x) && = \ph_1^{(0)}(x) + e\int\! d^4z\, G^R_1(x-z) \left[ 2
  A^\mu(z) \d_\mu\ph_2^{(0)}(z) + (\d A)(z)\ph_2^{(0)}(z)  - 2e\bP
  A^\mu(z)a_\mu^{(0)}(z)\right],\nn 
  \ph_2(x) && = \ph_2^{(0)}(x) -e \int\! d^4z\, G^R_2(x-z) \left[
  2A^\mu(z) \d_\mu \ph_1^{(0)}(z) + (\d A)(z)\ph_1^{(0)}(z) \right],
\end{eqnarray}
where the superscript zero denotes the solutions of the free EOM, and
$G^R$'s are the free retarded Green functions\footnote{Here the
  definition $KG^R=-\delta$ is used, where $K$ denotes the free
  kernel.}.

For the induced current (\ref{ind_curr}) we need the expectation
values of certain local products of the fluctuating fields. For
example, it directly follows from the above equations (with
$\exv{AB}_0\equiv\langle A^{(0)}B^{(0)}\rangle$), that
\begin{eqnarray}
  \exv{\ph_2(x)\d_\mu\ph_1(x)} =&& e\!\int\! d^4z \,\biggl\{
  \d_\mu^xG^R_1(x\!-\!z) \biggl[ 2 A^\nu(z)
  \exv{\ph_2(x)\d_\nu^z\ph_2(z)}_0 + (\d A)(z) \exv{
  \ph_2(x)\ph_2(z)}_0\biggr]\nn 
  - && G^R_2(x\!-\!z)\biggl[ 2 A^\nu(z)\exv{\d_\mu\ph_1(x)
  \d_\nu\ph_1(z)}_0 + (\d A)(z) \exv{\d_\mu\ph_1(x) \ph_1(z)}_0
  \biggr]\biggr\}.
\end{eqnarray}
We define the local products in a symmetric way (i.e.
$\frac12\exv{\ph_2(x)\d_\mu\ph_1(x) + \d_\mu\ph_1(x)\ph_2(x)}$) and
introduce
\begin{equation}
  \Delta(x-z) = \frac12\exv{\ph(x)\ph(z) + \ph(z)\ph(x)}_0.
\end{equation}
Then the above expression can be written in Fourier space as
\begin{equation}
  \exv{\ph_2\d_\mu\ph_1}(Q) = -e A^\nu(Q) \pint4p p_\mu
  (2p-Q)_\nu\left[ G^R_1(p) \Delta_2(Q-p) +
  G^R_2(Q-p)\Delta_1(p)\right].
\end{equation}
The evaluation of other expectation values goes along the same
line, finally giving
\begin{eqnarray}
  && e\exv{j_\mu}(Q) = -e^2 A^\nu(Q) \pint4p (2p-Q)_\mu (2p-Q)_\nu\left[
  G^R_1(p) \Delta_2(Q-p) + G^R_2(Q-p)\Delta_1(p)\right],\nn
  && 2e^2\bP \exv{a_\mu\ph_1}(Q)= -4e^2 m_W^2 A^\nu(Q) \pint4p \left[
  G^R_{\mu\nu}(p) \Delta_1(Q-p) + G^R_1(Q-p)\Delta_{\mu\nu}(p)\right],
  \nn&& e^2 A_\mu\exv{\ph^2} = e^2 A_\mu(Q) \pint4p
  \left[\Delta_1(p)+\Delta_2(p) \right].
\label{two_products}
\end{eqnarray}
Assuming that the free fluctuations are in thermal equilibrium,
the propagators can be related to the corresponding spectral functions
\cite{Landsmann}, which are the discontinuities of the free kernels of
eq.~\eqref{linopeom}. Introducing
\begin{equation}
  \Delta_{m^2}(p)= 2\pi\left(\frac12 + n(|p_0|)\right)
  \delta(p^2-m^2), \quad \textrm{and}\quad G^R_{m^2}(p) =
  \frac1{p^2-m^2 + i\ep p_0},
\end{equation}
where $n$ is the Bose--Einstein distribution, we have
\begin{equation}
  \Delta_1= \Delta_{m_H^2},\quad  \Delta_2 = \Delta_{m_G^2+\xi
  m_W^2},\quad \Delta_{\mu\nu} = -g_{\mu\nu} \Delta_{m_W^2} +
  \frac{p_\mu p_\nu}{m_W^2} \left( \Delta_{m_W^2} - \Delta_{\xi m_W^2}
  \right),
\end{equation}
and analogously for the corresponding $G^R$'s.

\section{High-temperature expansion}

The leading HTL term of $j^{ind}_\mu$ in the high-temperature
expansion comes from $e\exv{j_\mu}$ by neglecting all the masses
\cite{Rebhan95}.  Our aim is to calculate the first subleading term
proportional to $\bP^2$ instead of $T^2$ in the high-temperature
expansion.

We emphasize that $j^{ind}(Q)= j^{ind}(-Q)^*$, therefore all partial
contributions that are odd under hermitian $Q$-reflection can be
freely omitted.

A useful formula, which simplifies the calculations, is the following
\cite{JPPSz}:
\begin{eqnarray}
  I_F(Q,m^2,M^2) &=& \pint4p F(Q,p) \left(G^R_{m^2}(p)\Delta_{M^2}(Q-p) +
  G^R_{M^2}(Q-p)\Delta_{m^2}(p)\right) \nn &=& \pint4p F(Q,\frac
  Q2-p)\frac{\Delta_{m^2}(p-\frac Q2) - \Delta_{M^2}(p+\frac Q2)} {2pQ
  +m^2-M^2}.
\label{fund_eq}
\end{eqnarray}
The tensorial structures appearing in (\ref{two_products}) when cast
into the form (\ref{fund_eq}) imply the appearance in $F$ of the
$p$-dependent terms $p_\mu p_\nu, p_\mu Q_\nu +p_\nu Q_\mu$ and also
of terms independent of $p$.

\subsection{Even terms}

We start the discussion with the terms even under $p$-reflection and
introduce the notation $f(p)=(F(Q,Q/2-p)+F(Q,Q/2+p))/2$. Here the
$m^2-M^2$ term in the denominator can be neglected since in the mass
expansion 
\begin{equation}
  \pint4p f(p) (M^2-m^2) \frac{\Delta_0(p-\frac Q2) - \Delta_0(p+\frac
  Q2)}{(2pQ)^2} =0
\end{equation}
because of the odd $p\to-p$ behaviour of the integrand. Then we only
need to expand the difference of $\Delta$'s of the numerator.  The
numerator can be expanded with respect to $Q$, when low-momentum mean
fields are considered. The leading term of the gradient expansion
gives zero, again because of the $p$-odd integrand.

The first non-zero contribution is therefore
\begin{equation}
  I_F(m^2,M^2) = -\frac{Q_\rho}2 \pint4p \frac{f(p)}{2pQ} \frac\d{\d
  p_\rho}(\Delta_{m^2}(p) + \Delta_{M^2}(p)).
\end{equation}

Now, we treat separately the two actual cases: $f=p_\mu p_\nu$ and
$f=\,$constant. The $f=p_\mu p_\nu$ case contributes to the full
expression of $j^{ind}_\mu (Q)$
\begin{equation}
  2e^2 A^\nu(Q) Q^\rho \pint4p \frac{p_\mu p_\nu}{2pQ}\frac\d{\d
  p^\rho}(\Delta_1+\Delta_2 +\Delta_{m_W^2} -\Delta_{\xi m_W^2}).
\label{pp_contr}
\end{equation}
We introduce the field-strength tensor by $A^\nu Q^\rho=-i F^{\nu\rho}
+ A^\rho Q^\nu$. The local term proportional to $A$ reads as
\begin{equation}
  e^2 A^\rho(Q) \pint4p p_\mu \frac\d{\d
  p^\rho}(\Delta_1+\Delta_2 +\Delta_{m_W^2} -\Delta_{\xi m_W^2}).
\end{equation}
After partial integration the first two terms cancel with $e^2
A_\mu\exv{\ph^2}$ in the induced current. What remains is a
contribution vanishing at zero $m_W^2$. In the mass expansion they
give
\begin{equation}
  j_\mu^{local,1} = -e^2 (1-\xi) m_W^2 A_\mu(Q) \pint4p
  \frac{\d\Delta_0}{\d\p^2}.
\label{coupling1}
\end{equation}
For each $\Delta_{m^2}$ the field strength contribution of
(\ref{pp_contr}) is rewritten with the help of the relation
\begin{equation}
  s^\rho\frac{\d\Delta}{\d p^\rho} = -2ps \frac{\d\Delta}{\d\p^2} +
  s_0 \frac{dn(|p_0|)}{dp^0} 2\pi \delta(p^2-m^2)
\label{aux_rel}
\end{equation}
in the form
\begin{equation}
  -2ie^2 \pint4p \left( -\frac{p_\mu}{pQ} p_\nu p_\rho F^{\nu\rho}(Q)
   \frac{\d\Delta_{m^2}}{\d\p^2} + F^{\nu 0}(Q) \frac{p_\mu p_\nu}{2pQ}
   \frac{dn(|p_0|)}{dp_0} 2\pi\delta(p^2-m^2)\right).
\end{equation}
The first term drops out because of the antisymmetry of $F$. In the
second term we perform the mass expansion. After adding the different
contributions we find
\begin{equation}
  j_\mu^{(1)} = 4e^2(-iF^{\nu 0}) \pint4p \frac{p_\mu p_\nu}{2pQ}
   \frac{dn(|p_0|)}{dp_0} \left[ 2\pi\delta(p^2) - \frac{m_H^2+m_G^2 +
   m_W^2}2 2\pi\delta'(p^2)\right]\equiv \Pi_{\mu\nu}A^\nu.
\label{jmu1}
\end{equation}
The first term of this expression is the usual HTL contribution, the
second is a mass correction to it. In the mass correction, originally,
there was $m_1^2+m_2^2+m_W^2-\xi m_W^2$, but because of
$m_2^2=m_G^2+\xi m_W^2$ the $\xi$ dependence drops out.

The $f=\,$constant contribution to the induced current appears as
\begin{equation}
  -4e^2m_W^2A_\mu(Q) Q^\rho\pint4p \frac1{2pQ} \frac{\d\Delta_0}{\d
   p^\rho}.
\end{equation}
Since it is already proportional to $m_W^2$ we have dropped the mass
dependence of the integral. Using once more the relation
(\ref{aux_rel}) we write for it
\begin{equation}
  -4e^2 m_W^2 A_\mu(Q) \pint4p \frac1{2pQ} \left(
   -2pQ\frac{\d\Delta_0}{\d \p^2} + Q_0 \frac{dn(|p_0|)}{dp_0}
   2\pi\delta(p^2)\right).
\end{equation}
The first term is again local:
\begin{equation}
  j_\mu^{local,2} = 4e^2 m_W^2 A_\mu(Q) \pint4p \frac{\d\Delta_0}{\d \p^2}.
\label{coupling2}
\end{equation}
The second term reads as
\begin{equation}
  j_\mu^{(2)} = -4e^2 m_W^2 Q_0 A_\mu \pint4p \frac1{2pQ}
  \frac{dn(|p_0|)}{dp_0} 2\pi\delta(p^2).
\label{jmu2}
\end{equation}

\subsection{Odd contributions}

Odd contributions come exclusively from $\exv{a_\mu\ph_1}$. 
Using \eqref{fund_eq} and denoting the current by $j_\mu^{(3)}$ we find
\begin{equation}
  j_\mu^{(3)} = 2e^2A^\nu \pint4p (p_\mu Q_\nu + p_\mu Q_\nu)
  \frac{\Delta_{m_W^2}(p-\frac Q2) - \Delta_1(p+\frac Q2)} {2pQ
  +m_W^2-m_1^2} - \{m_W^2 \to \xi m_W^2\}.
\end{equation}
Finally, with the help of \eqref{aux_rel}, performing the mass and
external momentum expansion according to the method followed in the
previous subsection we arrive at
\begin{equation}
  j_\mu^{(3)} = 2e^2(1-\xi)m_W^2 Q_0 A^\nu \pint4p \frac{p_\mu Q_\nu +
  p_\mu Q_\nu} {(2pQ)^2} \frac{dn(|p_0|)}{dp_0} 2\pi\delta(p^2)\equiv
  Z_{\mu\nu} A^\nu.
\label{jmu3}
\end{equation}

\subsection{Linear response and physical modes}

The full induced current is the sum of the different parts coming from
\eqref{coupling1}, \eqref{jmu1}, \eqref{coupling2}, \eqref{jmu2} and
\eqref{jmu3}:
\begin{equation}
  j_\mu^{ind} = j_\mu^{local} + j_\mu^{(1)} + j_\mu^{(2)} + j_\mu^{(3)},
\end{equation}
where $j_\mu^{local}=j_\mu^{local,1}+j_\mu^{local,2}$.

Here $j_\mu^{(1)}$ and $j_\mu^{(2)}$ are $\xi$-independent, while
$j_\mu^{local}$ and $j_\mu^{(3)}$ depend on the gauge fixing.  For the
physical characterization of the system (for instance, damping rates)
we have to find the independently evolving modes. In order to do this
we decompose the polarization tensor in the tensor basis, appropriate
for finite-temperature studies \cite{Buchmuller94}:
\begin{equation}
  P^T_{\mu\nu} = - g_{\mu i}(\delta_{ij} - \hat Q_i\hat Q_j) g_{\nu
  j},\quad P^L_{\mu\nu} = - \frac{Q^2}{q^2} u_\mu^T u_\nu^T,\quad
  P^G_{\mu\nu} = \frac{Q_\mu Q_\nu}{Q^2},\quad S_{\mu\nu} =
  \frac1q\left(Q_\mu u_\nu^T + u_\mu^T Q_\nu\right),
\label{projectors}
\end{equation}
where $Q=(q_0,\q)$, $\q^2=q^2$ and $u_\mu^T =g_{\mu0} - Q_\mu q_0
/Q^2$ was used.

Since $\Pi_{\mu\nu}$ of (\ref{jmu1}) is transverse
($Q^\mu\Pi_{\mu\nu}=0$), it is the combination of $P^T$ and $P^L$.
Introducing $\Pi_L=\Pi_{00}$ and $\Pi_T = 1/2P^T_{\mu\nu}\Pi^{\mu\nu}$
we find
\begin{equation}
  \Pi = -\frac {Q^2}{q^2} \Pi_LP^L + \Pi_T P^T.
\end{equation}
The $Z_{\mu\nu}$ term in (\ref{jmu3}) can be written as a combination
of $P^G$ and $S$. Introducing $Z_G=Z_\mu^\mu$ and $Z_S=(q_0^2 Z_G -
Q^2 Z_{00})/(qq_0)$ we find
\begin{equation}
  Z = Z_GP^G + Z_S S.
\end{equation}
Using the completeness relation $P^L+P^T+P^G=g$ the Fourier transform
of the EOM (\ref{gauge_eq}) has the following form:
\begin{equation}
  \left[(Q^2-R-\Pi_T)P^T + \left(Q^2-R+\frac{Q^2}{q^2} \Pi_L\right)
  P^L + \left(\frac1\xi Q^2 -R-Z_G\right)P^G - Z_SS\right]A =0,
\end{equation}
with $R$ defined from $ m_W^2 A_\mu +j_\mu^{local} + j_\mu^{(2)}
\equiv R A_\mu .$

Since $S$ is not a projector, but mixes the subspaces belonging to
$P^L$ and $P^G$, in this two-dimensional subspace the inverse
propagator matrix still has to be diagonalized:
\begin{equation}
  \left(\begin{array}[c]{cc} Q^2-R+\frac{Q^2}{q^2} \Pi_L &
  -Z_S\cr Z_S &\frac1\xi Q^2 -R-Z_G\cr \end{array}\right).
\end{equation}
The eigenvalues should be found with one-loop accuracy, which means
that terms proportional to the square of one-loop corrections (i.e.
$\Pi^2_L$, $Z_G^2$ and $Z_S^2$) should be neglected. This, however,
implies that the eigenvalues are the diagonal entries. With
appropriately rotated projectors in the $(P^G,P^L)$ plane, we can
therefore write
\begin{equation}
  \left[(Q^2-R-\Pi_T)P^T + \left(Q^2-R+\frac{Q^2}{q^2} \Pi_L\right)
  \tilde P^L + \left(\frac1\xi Q^2 -R-Z_G\right)\tilde P^G\right]A =0,
\label{gauge_phys_eq}
\end{equation}
As expected, the transverse modes plus the $\tilde P_L$ mode, which
might be called longitudinal, can be made independent of the
gauge-fixing parameter (see the discussion on the renormalization of
$R$ below).

\subsection{The infrared separation scale}

The $p$-integrals in all terms of $j_\mu^{ind}$ are factorized into a
radial and an angular integral. It is well known that the HTL term
describes the dynamical screening of the electric fluctuations below
the Debye scale, an infrared separation scale (``IR cut-off'') has to
be therefore introduced into the radial integration at $p=M=C_M\times
eT$. The effective equations one arrives at in this way are to be used
for the modes $p\le M$.

Applying this cut-off to the integrals appearing in the expressions of
$R(Q)$ and $\Pi_{\mu\nu}$, one finds
\begin{eqnarray}
  && R = m_W^2 \left[ 1 + \frac{3+\xi}{8\pi^2}e^2 \ln\frac{\kappa
  T}\Lambda + \frac{e^2T}{2\pi^2 M} \frac{q_0}q \ln\frac{q_0+q}{q_0-q}
  \right],\nn
  &&\Pi_L = m_D^2 \left(1-\frac{q_0}{2q}\ln\frac{q_0+q}{q_0-q}\right)
  + \frac{q^2}{Q^2} \frac{e^2 T}{4\pi^2 M}(m_W^2+m_G^2+m_H^2), \nn
  &&\Pi_T = m_D^2 \frac{q_0}{2q}\left[\frac{q_0}q-\frac{q_0^2-Q^2}{2q^2}
    \ln\frac{q_0+q}{q_0-q}\right] + \frac{e^2 T}{4\pi^2M}
  (m_W^2+m_G^2+m_H^2) \frac{q_0}q\ln\frac{q_0+q}{q_0-q},
\end{eqnarray}
with $\kappa=2\pi \exp (-\gamma_E)$ and $m_D^2=e^2T^2/3 -
e^2MT/\pi^2$.  The logarithmical UV divergence in $R$ can be absorbed
into the $e^2$ renormalization. With appropriate renormalization scale
$\mu=\kappa T$ the $\xi$ dependence can be made to vanish.

The intermediate IR scale $M$ contributes a term to the Debye mass,
which depends linearly on $M$. It will be cancelled by the linear
$M$-dependence of the self-energies, which shows up in the (one-loop)
solution of the classical effective EOM (\ref{gauge_eq})
\cite{bodeker95,arnold97,Hindmarsh99,aarts00}.

The mass corrections in (\ref{gauge_phys_eq}), proportional to $\bP^2
A^\nu$, should be interpreted as coming from an induced (non-local)
$\Phi^2 A^2$ vertex correction. In the subsequent classical time
evolution it contributes to the self-energies of both fields.  When
(perturbatively) combined with the $\sim MT$ terms coming from the
tadpoles, it results in a finite contribution of order $T^2$,
independent of the exact choice of the coefficient $C_M$ in the
expression of $M$. Therefore, these terms can by no means be
neglected.

\section{Conclusions}

In this letter we have investigated the induced current in the
effective EOM of the gauge field in the Abelian Higgs model.  Beyond
the well-known HTL term, we now determined also the subleading
$\sim\Phi^2 A_\mu$ corrections in adiabatic approximation
($\Phi=$~constant). The most important property of the corrections is
that, just as the leading HTL term, in the high-temperature
expansion for the physical degrees of freedom it is independent of the
gauge-fixing procedure (after appropriate renormalization).

For consistency we have applied an IR cut-off $M=C_M\times eT$ to the
fluctuations, and the effective EOM is valid below this scale. The
subsequent 3D time evolution will be insensitive to the accurate
choice of $M$.  Partly it is cancelled by the 3D
(Rayleigh--Jeans-type) divergences, partly it yields also non-zero,
$M$-independent contributions, when the gauge--scalar vertex
corrections are combined with the 3D would-be divergences.

We shortly discuss here the numerical implementation of the effective
gauge field equation. The most convenient gauge fixing seems to be the
Landau-gauge, where the $\xi$-dependent mode does not propagate. Then
$j_\mu^{(3)}$ can be left out of the discussion and the effective
equations of motion are to be solved under the constraint
$\partial_\mu A^\mu =0$. It can be implemented, for example, by
solving the equations for $A_i$ and computing $A_0$ from the
constraint.

The non-local induced currents are written in local form with the help
of auxiliary fields. With the well-known form of the leading HTL
current, we can write in this form also the corrections due to the
non-zero scalar background:
\begin{eqnarray}
&& j_i^{(1)}(x)=m_D^2\int\frac{d\Omega_v}{4\pi}\left(v_iv_j -
  \frac{e^2T}{2\pi^2M}\frac{m_W^2+m_G^2+m_H^2}{m_D^2}\delta_{ij}\right)
  W_j^{(1)}(x,{\bf v}),\nn
&& j_i^{(2)}(x)=\frac{e^2m_W^2T}{\pi^2 M}\int \frac{d\Omega_v}{4\pi}
W_i^{(2)}(x,{\bf v}),
\end{eqnarray}
where the auxiliary fields satisfy
\begin{equation}
(\d_0-\mbox{\boldmath $v\d$}) W_i^{(1)} = F_{0i},\quad
  (\d_0-\mbox{\boldmath $v\d$}) W_i^{(2)} = \d_0 A_i.
\end{equation}
In the derivation of the expression of $j_\mu^{(1)}$ we have performed
a partial $p$-integration in the part of (\ref{jmu1}) proportional to
$\delta'(p^2)$, and used the fact that, to the accuracy of our
calculation, $Q_\nu F^{\nu\mu}(Q)=0$ can be exploited in the
expression of the induced current.

The equations for the scalar field receive mainly local corrections
(renormalization and $T$-dependence of the couplings in the classical
equations). However, for consistency also the $\Phi$-derivative of the
$\Phi^2$--$A^2$ vertex correction established in the present
calculation should be introduced.

The logics of the derivation followed in the Abelian Higgs model seem
to be robust enough for us to attempt its generalization to the
non-Abelian case, which will be the subject of a future study.  A
numerical study of the corrected EOMs can lead us to a deeper
understanding of the reliability of our present views on the
high-temperature Higgs models, especially where the sphaleron rate is
concerned. We will also learn in more detail of the non-equilibrium
aspects of the cosmological onset of the Higgs regime.

\newpage
{\bf Acknowledgement}

The authors acknowledge fruitful discussions with P. Petreczky and H.
J. de Vega in the early stages of this work. A.J. thanks D. Litim, C.
Manuel, G. Moore, A. Rebhan and M.~Thoma, for valuable comments and
discussions. This research was supported by the Hungarian Science Fund
(OTKA).


\begin{thebibliography}{99}
\bibitem{Garcia99} J. Garcia-Bellido, D.Yu. Grigoriev, A. Kusenko and
M. Shaposhnikov, Phys. Rev. {\bf D60} (1999) 123504
\bibitem{Krauss99}L.M. Krauss and M. Trodden, Phys. Rev. Lett. {\bf
83} (1999) 1502
\bibitem{Rajantie00} A. Rajantie, P.M. Saffin and E.J. Copeland,
DAMTP-2000-134, hep-ph/0012097
\bibitem{Moore01}G.D. Moore and K. Rummukainen, Phys. Rev. {\bf D63}
(2001) 045002
\bibitem{Moore00}G.D. Moore, Phys. Rev. {\bf D62} (2000) 085011
\bibitem{Ambjorn95}J. Ambjorn and A. Krasnitz, Phys. Lett. {\bf B362}
(1995) 97
\bibitem{Braaten90} E. Braaten and R.D. Pisarski, Nucl. Phys. {\bf
    B334} (1990) 569
\bibitem{Frenkel90} J. Frenkel and J.C. Taylor, Nucl. Phys. {\bf B334}
(1990) 199
\bibitem{Blaizot94} J.-P. Blaizot and E. Iancu, Nucl. Phys. {\bf B417}
(1994) 608 
\bibitem{Hindmarsh99} A. Rajantie and M. Hindmarsh, Phys. Rev. {\bf
D60} (1999) 096001
\bibitem{Kajantie97} K. Kajantie, M. Laine, K. Rummukainen and
M. Shaposhnikov, Phys. Rev. Lett. {\bf 77} (1996) 2887
\bibitem{Karsch97} F. Karsch, A. Patk{\'o}s, T. Neuhaus and A. Rank,
Nucl. Phys. Proc. Suppl. {\bf 53} (1997) 623
\bibitem{Guertler97} M. G\"urtler, E.M. Ilgenfritz and A. Schiller,
Phys. Rev. {\bf D57} (1997) 3888
\bibitem{Csikor99} F. Csikor, Z. Fodor and J. Heitger,
Phys. Rev. Lett. {\bf 82} (1999) 21
\bibitem{Rebhan95} U. Kraemmer, A. K. Rebhan and H. Schulz,
  Ann. Phys. (NY) {\bf 238} (1995) 286.
\bibitem{Baacke97} J. Baacke, K. Heitmann and C. P{\"a}tzold,
Phys. Rev. {\bf D55} (1997) 7815
\bibitem{Baacke99} J. Baacke and K. Heitmann, Phys. Rev. {\bf D60}
(1999) 105037
\bibitem{Heitmann01} K. Heitmann, Gauge fields out of equilibrium: a
gauge invariant formulation and the Coulomb gauge, DO-TH-01/02,
LA-01-356, hep-ph/0101281 
\bibitem{LeBellac96} M. Le Bellac, {\it Thermal Field Theory}, Cambridge
Univ. Press, 1996
\bibitem{Boyanovsky98} D. Boyanovsky, D. Cormier, H.J. de Vega,
R. Holman and S.P. Kumar, hep-ph/9801453, in Procs. of 6th Erice
Chalonge Course on Astrofundamental Physics, N. Sanchez and
A.~Zichichi eds., Kluwer, 1998
\bibitem{Heinz86} U. Heinz, Ann. of Phys. (N.Y.) {\bf 168} (1986) 148
\bibitem{Mrovczinsky90} P. Danielewicz and H. Mrowczynski,
Nucl. Phys. {\bf B342} (1990) 345
\bibitem{Jakovac01} A. Jakov\'ac (in preparation)
\bibitem{Landsmann} N. P. Landsmann and Ch. G. van Weert,
  Phys. Rep. {\bf 145} (1987) 143
\bibitem{JPPSz} A. Jakov\'ac, A. Patk{\'o}s, P. Petreczky and
  Zs. Sz{\'e}p, Phys. Rev. {\bf D61} (2000) 025006
\bibitem{JP97} A. Jakov\'ac and A. Patk{\'o}s, Nucl. Phys. {\bf B494}
  (1997) 54
\bibitem{Buchmuller94} W. Buchm\"uller, Z. Fodor, T. Helbig and
  D. Walliser, Ann. Phys. (N.Y.) {\bf 234} (1994) 260
\bibitem{bodeker95} D. B{\"o}deker, L. McLerran and A. Smilga,
  Phys. Rev. {\bf D52} (1995) 4675
\bibitem{arnold97} P. Arnold, D. Son and L.G. Yaffe, Phys. Rev. {\bf
    D55} (1997) 6264
\bibitem{aarts00} G. Aarts, B-J. Nauta and Ch. G. van Weert,
  Phys. Rev. {\bf D61} (2000) 105002
\end{thebibliography}
\end{document}